\documentclass[onecolumn,amsmath,amssymb]{IEEEtran}
\usepackage{graphicx,float}
\usepackage{color}
\usepackage{amsmath}
\usepackage{amssymb}
\usepackage[abs]{overpic}     

\begin{document}

\title{Ultrasound imaging of the human body with three dimensional full-wave nonlinear acoustics.\\ Part 1: simulations methods.}
\author{Gianmarco~Pinton$^1$\\$^1$Joint Dept. of Biomedical Engineering, University of North Carolina at Chapel Hill and North Carolina State University\\}
\thanks{Further author information: \\Send correspondence to G. Pinton: E-mail: gia@email.unc.edu}
 
\markboth{}{Pinton: }
\maketitle

\newcommand{\invivo}{\emph{in vivo}} 
\newcommand{\exvivo}{\emph{ex vivo}}
\newcommand{\insilico}{\emph{in silico}}
\newcommand{\insitu}{\emph{in situ}}



\begin{abstract}
Simulations of three dimensional ultrasound propagation in heterogeneous media are computationally intensive due to the combined constraints arising from the large size of the domain, which is on the order of hundreds of wavelengths, and the small size of scatterers, which can be much smaller than a wavelength. For this reason three dimensional ultrasound imaging simulations are currently based on models that simplify the propagation physics. Here the full three dimensional wave physics is simulated with finite differences to generate ultrasound images of the human body based directly on the first principles of propagation and backscattering. The Visible Human project, a 3D data set of the human body that was generated with photographs of 0.33 mm cryosections, is converted into 3D acoustical maps. A full-wave nonlinear acoustic simulation tool is used to propagate ultrasound into the liver with a 2D transcostal ultrasound array in a $93 \times 39 \times 22$ mm domain with $6\times10^8$ points. Imaging metrics, based on the beamplots, root-mean-square phase aberration, spatial coherence lengths, and contrast-to-noise ratio are used to characterize the image quality. It is shown that the harmonic image quality is better than the fundamental image quality due, in part, to a narrower beam profile. The root-mean-square estimate of aberration after propagation through the simulated body wall is shown to be low (23.4 ns), which is consistent with previous reports of aberration measured experimentally in a human body wall. The spatial coherence measured at the transducer surface indicates that a transducer array element size of $<0.81 \lambda$ would be required to fully sample the acoustic field. These first simulated three dimensional ultrasound images based directly on propagation physics provide a platform to investigate the sources of image degradation in three dimensions. A detailed charaterization of these sources of image degradation, including reverberation clutter, are included in Part II of this paper.

\end{abstract}

\section{Introduction}

The use of realistic simulations of ultrasound imaging in the body has
a wide variety of applications such as beamforming optimization,
transducer design, estimating the pressures in the human body for
safety considerations, and many others. The generation of an
ultrasound image of the soft tissue in the body relies on the physics
of acoustic wave propagation: diffraction, reflection, scattering,
frequency dependent attenuation, and nonlinearity. At the most
fundamental level an ultrasound image relies on an acoustic wave
propagating to a target, reflecting, and then propagating back to the
transducer.

However relying directly on propagation physics to simulate an
ultrasonic image is inherently computationally costly. This
is due to two fundamental physical scales. First, there are a large
number of propagation wavelengths ($\sim 100 \lambda$) and the
simulation domain must therefore encompass these large
distances. Second, the ultrasonic pulse is backscattered by
sub-wavelength structures ($<\lambda/10$). Therefore, to represent
both propagation and backscattering, the large spatial domain
determined by the first physical scale must be spatially sampled at
the relatively small scale determined by the second physical
scale. Furthermore simulating backscattering represents an additional
numerical challenge because the low amplitude reflections require a
high dynamic range, significant accuracy at material interfaces, and
suppression of unwanted reflections from the simulation boundaries.

One approach to simulating ultrasound images is to make approximations
that reduce the physics to systems that have a lower computational
cost. This popular due to its effectiveness and there are too many
examples to list fully. The Field II simulation tool, is perhaps the
best known example, and it simplifies certain aspects of wave
propagation by assuming linear propagation and by calculating a
spatial impulse response~\cite{Jensen1996}. This eliminates the need
to simulate wave propagation directly by using a convolution
approach. Consequently it can accurately model a wide range of linear
multi-element transducers with a variety of apodization, focusing, and
excitation configurations with comparatively low computational
cost. However, by not simulating wave propagation directly, effects
such as mulitple scattering, reverberation, and dsitrubuted
aberration, which can be determining factors of ultrasound image
quality, cannot be modeled.


There are a number of simulation tools that model nonlinear ultrasound
propagation. Angular spectrum methods or retarded time methods, for
example, typically model propagation by changing the frame of
reference so that the simulation domain follows the wave as it
travels~\cite{Christopher1991,Zemp2003,
  dagrau2011acoustic,yuldashev2011simulation}. This yields substantial
numerical benefits, especially for long propagation paths and for
capturing large amplitudes and shock wave development. These
advantages explain why these methods have been so successful in
modeling therapeutic acoustic fields. However they cannot model
multiple reflections and backpropagation, which are necessary for
imaging.  Ultrasonic propagation through fine scale heterogeneities
has been simulated previously with a finite difference time domain
(FDTD) solution of the 2D and 3D linear wave equation
\cite{Waag1997,Mast2002}. Other simulation tools also exist that can
model nonlinear ultrasound propagation based on k-space methods
~\cite{treeby2010k}. These numerical tools can model the fundamental
wave physics of backpropagation, nonlinearity, and
attenuation. However these simulations have not been used to generate
ultrasound images based on the first principles of propagation and
reflection.

This direct propagation approach and the full three dimensional
nonlinear propagation wave physics is used here to generate highly
realistic ultrasound images. These simulatons are based on a
previously developed a numerical solution tool that solves the
full-wave equation and which will be referred to as
``Fullwave''~\cite{pinton2009heterogeneous}. In addition to simulating
the nonlinear propagation of waves it describes arbitrary frequency
dependent attenuation and variations in density.  This simulation tool
has been previously used to generate ultrasound images, to study the
sources of image
degradation~\cite{pinton2011sources,pinton2011effects} and to
understand the principles behind new imaging methods, such as short
lag spatial coherence imaging~\cite{pinton2014spatial}.  It has also
been used to simulate how elements that are blocked by ribs can
degrade the image quallity~\cite{jakovljevic2017blocked-a}.  Although
the capability to simulate in three dimensions with Fullwave has
existed since its inception and although it has been used extensively
in three dimensional therapuetic
applications~\cite{pinton2011cavitation,pinton2011effects}, it's use
for imaging in three dimensional domains has not been previously
described.

The objective of this paper is to model nonlinear acoustic propagation
and backscattering to generate physically realistic three dimensional
ultrasound images in the human body. One of the challenges of using a
direct propagation approach is finding an appropriate data set that
accurately represents the acoustical properties of the human body in
three dimensions. To achieve this goal The National Library of
Medicine's Visible Human data set is used. This publically available
data set was generated for a male and a female cadaver whose
cryosections were photographed and
digitized~\cite{spitzer1996visible}. This data set also includes MRI
and CT (but not ultrasound) scans. We present an image processing
method to convert the optical data set into maps of the acoustical
properties for an intercostal imaging scenario. These acoustical maps
could be used in any acoustical wave propagation solver. Here they are
used with the Fullwave simulation tool to generate propagation-based
fundamental and harmonic B-mode ultrasound images. The simulation
generates images in the same way that an ultrasound scanner would
generate images. Sound is emitted from a transducer, it propagates in
the tissue, and it is reflected. The backscattered sound is measured
at the transducer surface and then filtered and beamformed to generate
fundamental and harmonic B-mode images. Several imaging properties are
studied, in particular the beamplots, the spatial coherence at the
transducer surface, and the lesion detectability for imaging at the
fundamental and harmonic frequencies. In part II of this two-part
paper the sources of image degradation in intercostal imaging are
examined in detail. To determine the effects of the ribs on beamplots,
phase aberration, and reverberation clutter three configurations are
simulated: with the ribs in place, with the ribs removed, and with the
ribs placed closer together. This configuration provides a imaging
scenario that could not be performed {\it in vivo} and it uses the
imaging anlysis and simulation methods established here.

\section{Methods}
\subsection{Modeling equations}
A full description of the nonlinear full-wave equation and the finite
differences used to solve it can be found in the references
~\cite{Pinton2007} and these are summarized briefly here. The
nonlinear wave equation solved by Fullwave is based on the Westervelt
equation with the addition of relaxation mechanisms to account for the
non-classical attenuation observed in the soft tissue of the human
body:
\begin{equation}
0=\nabla^2 p - 
\frac{1}{c_0^2}\frac{\partial^2 p}{\partial t^2}+
\frac{\delta}{c_0^4}\frac{\partial^3 p}{\partial t^3}+
\frac{\beta}{\rho c_0^4}\frac{\partial^2 p^2}{\partial t^2}
\frac{1}{\rho}\nabla p \cdot \nabla \rho
-\sum_{m=1}^{v}\xi_m
\label{eqn:westervelt}
\end{equation}
\noindent The first two terms in Eq.~\ref{eqn:westervelt} represent
the linear wave equation, and the following three terms represent
thermoviscous diffusivity, nonlinearity, and variations in density.
The remaining term represents $v$ relaxation mechanisms, where $\xi_m$
satisfies the equation where $\xi_m$ satisfies the equation
\begin{equation}
\dot{\xi}_m+\omega_m\xi_m = a_m\omega_m\frac{\Delta c}{c_0}\nabla^2p
\label{eqn:relax}
\end{equation}
In these equations, $p$ is the acoustic pressure, $c_0$ and $\rho$ are
the equilibrium speed of sound and density, $\delta$ is the acoustic
diffusivity, $\alpha$ is the absorption coefficient, and the
coefficient $\beta$ is related to the nonlinearity parameter, $B/A$,
by the relationship $\beta=1+B/2A$.  The diffusivity can be expressed
as a function of the absorption coefficient with the equation
$\delta=2\alpha c_0^3/\omega^2$ (where $\omega$ is the angular
frequency).  The relaxation equation (Eq.~\ref{eqn:relax}) has $v$
peaks at characteristic frequencies $\omega_m$ with weight $a_m$ that
depend on the particular frequency dependent attenuation law being
modeled. This equation was solved using finite differences in the time
domain. Note that the the speed of sound, $c_0$, density, $\rho$,
attenuation, and nonlinearity, $\beta$, can all vary as a function of
space. Maps of these variables can be used to represent the
heterogeneous acoustical properties of the human body.

\subsection{Generation of three dimensional acoustic maps}

The photographic cryosections of the female Visible Human data set
were used because they have the highest resolution (0.33 mm slice
thickness). A number of image processing steps were performed to
transform the photographic data from the Visible Human data set to
three dimensional maps of the acoustical properties that can be used
generally used in simulations. The objective was to retain as much
detail as possible since ultrasound imaging is, by design, sensitive
to small echoes from fine material interfaces. For example, the thin
layers of connective tissue in abdominal fat provide an acoustically
rich structure that can generate aberration and reverberation.

A region was selected that contained two ribs, skin, fat, and muscle
layers, and the underlying liver. This corresponds to images 1432 to
1909 in the abdominal region of the Visible Human data set. A cropped
section is shown on the left of Fig.~\ref{fig:realization2} orthogonal
to the rib axis (top left) and parallel to the rib axis (bottom
left). The general image processing approach was to detect the
interface layers between different body wall regions, skin, fat,
muscle, and liver, and then within each region to detect different
tissue types. The image processing steps were implemented with Matlab
(Mathworks, Natick, MA, USA) and they are described in detail below:

\begin{figure}[H]
  \begin{center}
    \includegraphics[height=0.39\linewidth]{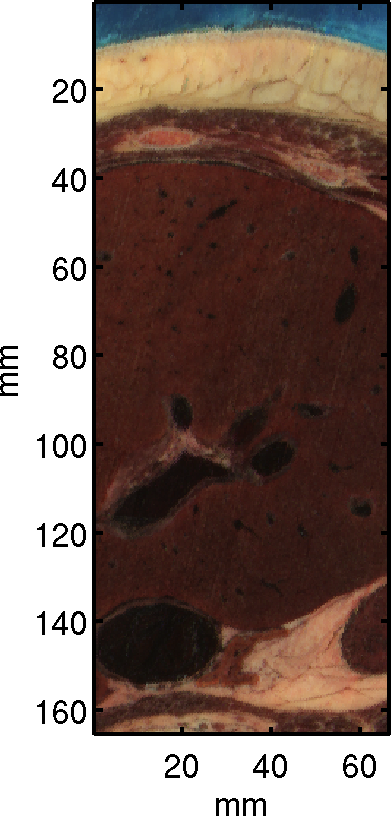}
    \includegraphics[height=0.39\linewidth]{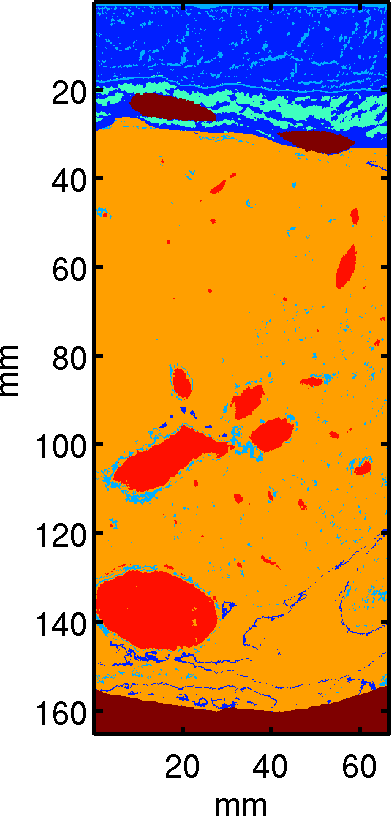}\\
    \includegraphics[height=0.39\linewidth]{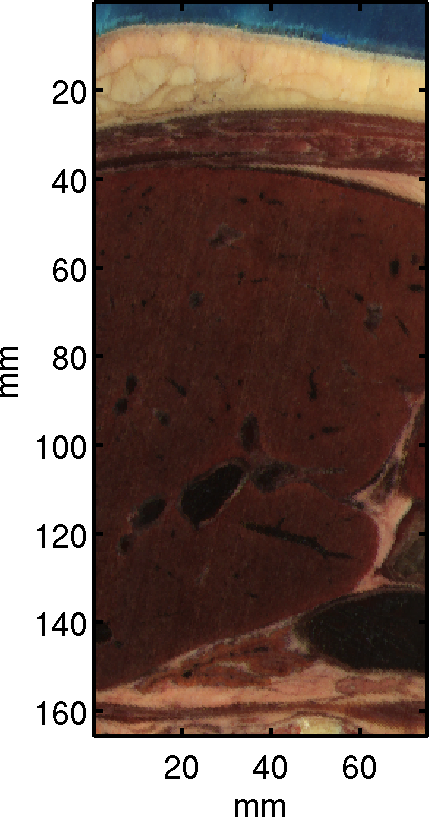}
    \includegraphics[height=0.39\linewidth]{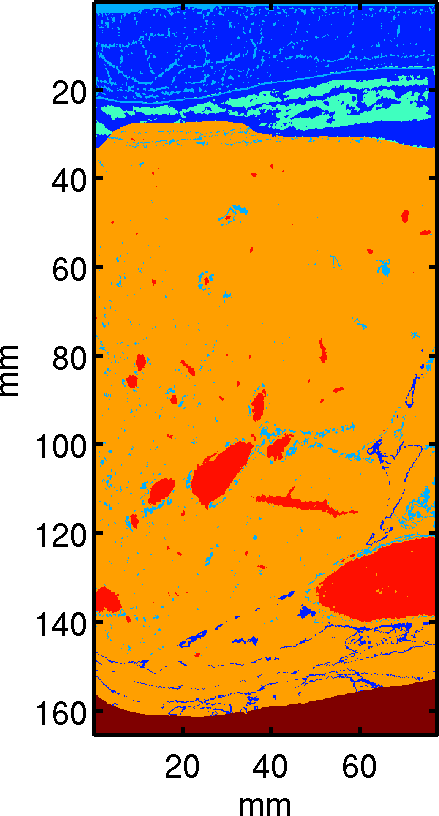}
 \end{center}
  \caption{Digitized tissue image in RGB format orthogonal to the rib axis (top left) and along the rib axis (bottom left). The index map that tags different tissue types after flattening the maps relative to the skin surface (right).}
  \label{fig:realization2}
\end{figure}

\subsubsection{Define the map coordinate system} The long axis of the transducer was oriented to be parallel to the ribs. The map coordinate system was then defined with respect to this axis. The vector in between the two ribs was determined manually by first selecting points in between the ribs, then by fitting a line to them with a standard least squares regression in three dimensions. With respect to the original coordinate system, defined in the abdominal Visible Human images, the centroid is given by (564.9375, 697.2500, 325.0000) and the direction cosines of the best fit line is given by (0.6202, 0.2403, 0.7467). To complete the definition of the coordinate system an additional point in the liver was used to define the center of the image, and it was manually selected to be (350, 967, 335). In summary, the lateral imaging plane is along the transcostal line and it intersects the point in the liver. The elevation plane is orthogonal to the transcostal line and it also intersects this point. This coordinate system was used as a reference for all other calculations. 
\subsubsection{Manually segment ribs} Since fat and bone/cartilage have a similar color on the optical images the ribs were segmented manually in each image. This segmentation data is too lengthy to reproduce here but can be provided upon request. 
\subsubsection{Interpolation and rotation} The imaging volume was linearly interpolated (from 0.330 mm to 0.167 mm) using two successive applications of the \texttt{interp2} Matlab function. Then the data set was rotated to the imaging axis coordinates. The rib segmentation coordinates underwent the same transformations. 
\subsubsection{Flatten body surface in two dimensions using an RGB threshold detection algorithm} The body surface is curved however the ultrasound transducer is pressed into the body during imaging. To flatten the body surface relative to the transducer the superficial skin layer was detected. This body curvature was defined by the boundary between the frozen outer gelatin block and the skin surface. It was calculated with a two dimensional RGB threshold detection filter. The upper RGB threshold was (95, 160, 180) and the lower threshold was (20, 50, 65). The 2D detected surface was then filtered with a $5 \times 5$ median filter. This curved surface was then used as a reference to flatten the imaging volume so that the outer skin layer was planar. This flattening can be seen by comparing the images on the left of Fig.~\ref{fig:realization2} to the images on the right. 
\subsubsection{Detect skin layer} The previous step effectively determined the superficial surface of the skin. To determine the depth of the skin layer an RGB threshold of (160, 130, 90) was used. This surface was filtered with a $20 \times 20$ median filter. 
\subsubsection{Detect fat layer and embedded connective tissue} The fat layer was detected with an RGB filter with a threshold of (160, 130, 90) for the skin/fat boundary (as described in the previous step), and a threshold of (130, 101, 56) for the fat/muscle boundary. Both detected surfaces were filtered with a $20 \times 20$ median filter. Embedded within the fat layer there is a fine connective tissue. The RGB values of this connective tissue vary significantly and intersect with the surrounding fat which makes threshold detection impossible. To determine the fine structures embedded in the tissue layers a gradient based edge detection filter was used. A cube of 1.155 mm per side was convolved with the green values in the image. This convolved image was then subtracted from the original image to obtain a gradient map that can be subsequently thresholded to detect the edges. Pixels with a gradient between -3 and -30 RGB/pixel were tagged as connective tissue. 
\subsubsection{Detect muscle layer and embedded fat} The muscle layer was detected with an RGB filter with a threshold of (99, 60, 50) for the muscle/liver boundary. Within the muscle layer the embedded fat was detected with a gradient based edge detection method on the green values using the same method described in the previous paragraph. The characteristic width of the filter was 5.115 mm and fat was detected for gradients between 0 and 50 RGB/pixel.
\subsubsection{Detect liver, and embedded connective tissue and blood vessels} Within the liver blood and vessels was detected for RGB values between (1, 1, 1) and (35, 30, 30). Connective tissue was detected for RGB values between (44, 26, 24) and (59, 40, 35). Fat was detected for RGB values between (102, 75, 43) and (136, 111, 95).

Once each tissue type was detected and tagged within the simulation
volume, the tissue map was converted into four maps of acoustical
properties: speed of sound, density, attenuation, and nonlinearity.

Their values for each tissue type are shown in
Table~\ref{tab:tissue}. In this imaging configuation the ribs were
modeled to have the same acoustic impedance as bone but with the speed
of sound and density reduced, instead of increased.  This was done to
have a larger temporal step size or equivalently CFL, which reduces
the calculation time and improves the numerical accuracy. The ampltude
of the reflections from bone surface were therefore accurately modeled
and propagation inside the bone was considered to be irrelevant since
the attenuation in bone is so large. Note that Fullwave has been used
extensively to propagate ultrasound through
bone~\cite{pinton2011cavitation,pinton2011effects} and that this
choice of an inverted impedance was motivated by optimization of the
time step rather than technical difficulties of propating in bone.

 \begin{table}[h]
\begin{center}
    \begin{tabular}{|c|c|c|c|c|}
        \hline
       Tissue  & Speed & Density & Attenuation  & Nonlinearity  \\
         ~ & m/s & kg/m$^3$ & dB/MHz/cm  & B/A  \\ \hline
       fat &  1479  &  937  & 0.4  & 9.6 \\
       liver&  1570  & 1064   & 0.5  & 7.6 \\
       muscle&  1566  & 1070  & 0.15  & 9 \\
       connective& 1613  &1120  & 0.5  & 8 \\
       blood& 1520  & 1000  & 0.005 & 5 \\
       bone & 800  &  550  & 5  & 5 \\
       \hline
      \end{tabular}
 \caption{Tissue properties}
 \label{tab:tissue}
\end{center}
 \end{table}

Since the optical resolution of the visible human data set is 0.33 mm,
small sub-wavelength scatterers aren't detectable. Nevertheless these
sub-resolution scatterers play an important acoustic role in the
imaging physics and in particular to obtain the correct speckle
statistics~\cite{wagner1983statistics}. Sub-resolution scatterers were
therefore added to the tissue maps to model the acoustical scattering
properties observed {\it in vivo}. A total of 20 scatterers with a 77
$\mu$m diameter were added per resolution volume (as calculated with a
2MHz F/2 aperture). This generates fully developed speckle. The
scatterers had a spatial distribution and random amplitude and with an
average impedance mismatch of 2.5\% relative to the background
tissue. An artificial anechoic lesion was placed at the 6.5 cm focus
by removing the scatterers in a spherical region within a 5 mm
radius. This anechoic lesion was used to characterize the imaging
properties.

The final speed of sound map for the simulation field is shown in
Fig~\ref{fig:realization2} in the lateral imaging plane (left),
elevation plan (center), and just under the transducer suface
(right). Not shown are the equivalent maps for attenuation,
nonlinearity, and density which are similar in appearance. These maps
include sub-resolution scatterers, which are not visible, and the
negative impedance mismatch for the ribs. The location of the anechoic
lesion is overlaid.

\begin{figure}[H]
  \begin{center}
\begin{overpic}[height=0.47\textwidth]{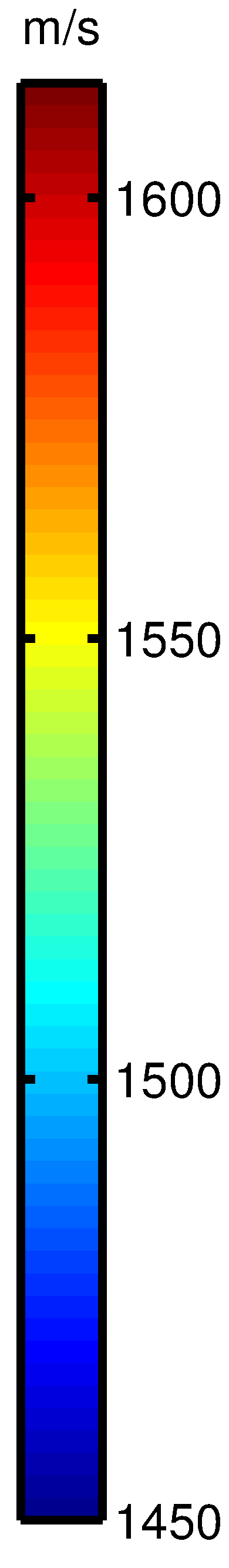}
\end{overpic}
\begin{overpic}[width=0.7\textwidth]{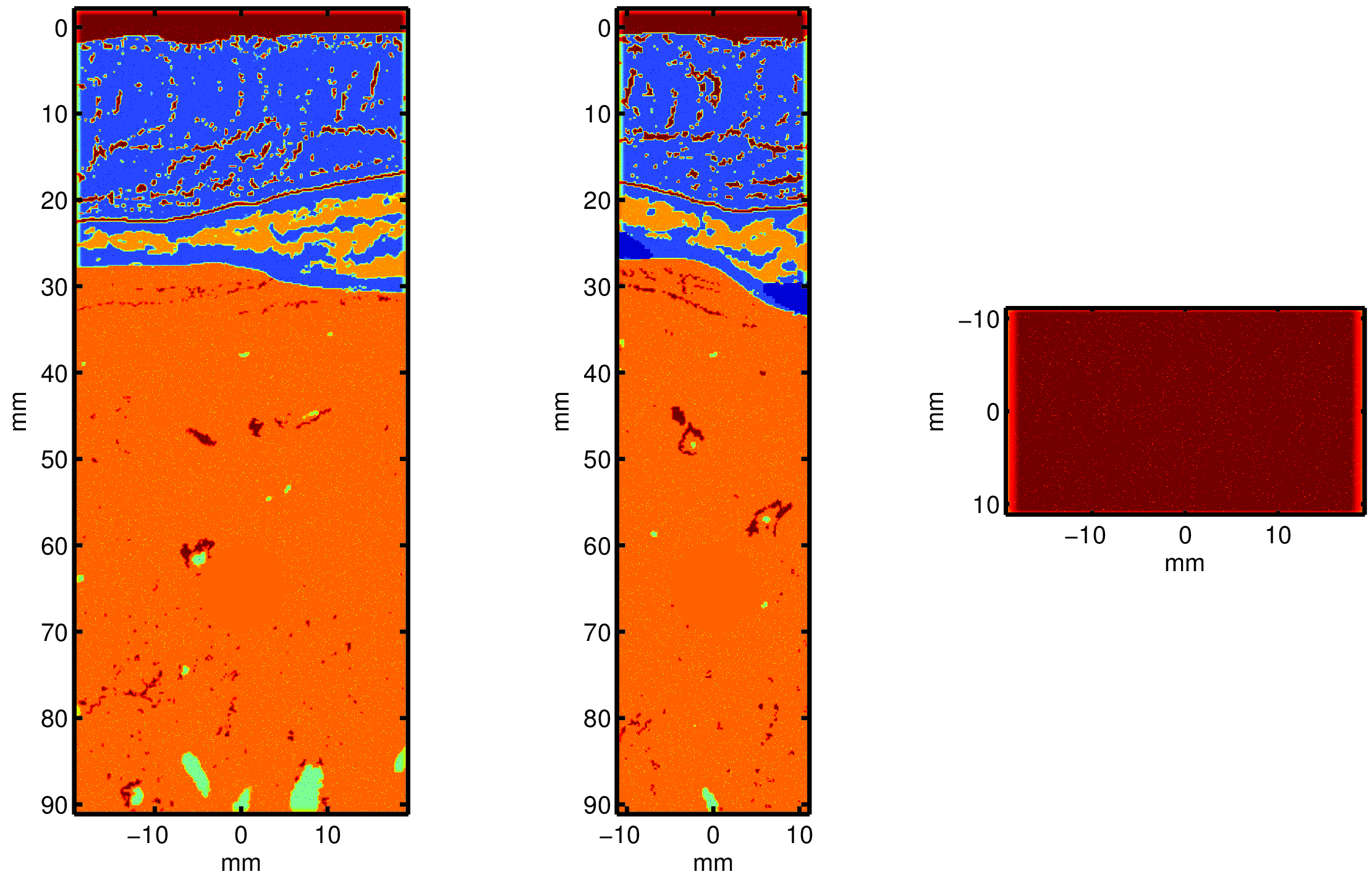}
\small
\put(194,206){$\leftarrow$skin}
\put(194,180){$\leftarrow$fat}
\put(194,160){$\leftarrow$connective}
\put(194,150){$\leftarrow$muscle}
\put(194,138){$\leftarrow$rib}
\put(194,70){$\leftarrow$liver}
\put(172,70){\color{white}{\circle{20}}}
\put(155,50){\textcolor{white}{anechoic}}
\put(160,40){\textcolor{white}{lesion}}
\put(59,70){\color{white}{\circle{20}}}
\put(43,50){\textcolor{white}{anechoic}}
\put(48,40){\textcolor{white}{lesion}}
\end{overpic}
  \end{center}
  \caption{The speed of sound map for the center transmit simulation
    field in the lateral imaging plane (left), elevation plan
    (center), and just under the transducer suface (right). Note the
    inclusion of sub-resolution scatteres and the negative impedance
    mismatch for the ribs. The location of the anechoic lesion is
    overlaid.}
  \label{fig:realization2}
\end{figure}
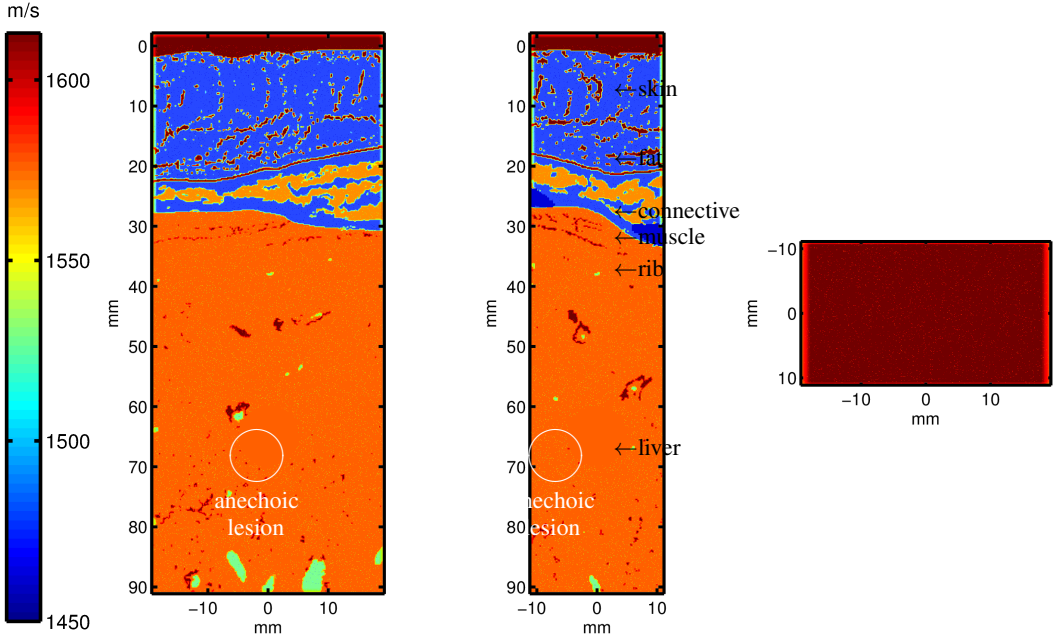

\subsection{Imaging configuration and simulation parameters\label{sec:transmit}}

An aperture meant to represent a general transcostal imaging
transdcuer was used to transmit and receive ultrasound. These
transmit-receive simulation elements were placed at the skin surface
of the heterogeneous acoustic tissue maps. The transducer was modeled
as a 3.25 $\times$ 1.625 cm 2D array, which is consistent with the
sizes typically chosen for intercostal liver
imaging~\cite{szabo2013ultrasound}.  A a 2.5 cycle, 2 MHz, 0.2 MPa
pulse focused at 65 mm was emitted without apodization.  These
transmit parameters are summarized in Table~\ref{tab:transducer}.  The
ultrasound imaging sequence consisted of 5 transmit receive
events. For each event the transducer was shifted laterally by a
beamwidth calculated as 1.4 $\lambda z/d$. This shift is the
equivalent of walking the aperture, or of mechanical
translation. Other transmit options, such as beam steering, could have
been simulated but this configuration was chosen for its
simplicity. Parallel receive beamforming methods were used on the
received data and they are described subsequently in sec
\ref{sec:image}.

\begin{table}[H]
\begin{center}
     \begin{tabular}{|c|c|}
        \hline
       Transducer  & Value \\ 
       Parameter & \\ \hline
       Center freq. & 2.0 MHz\\ 
       Width & 3.25 cm \\
       Elevation & 1.625 cm \\ 
       Number of cycles & 2.5 \\ 
       Tx pressure & 0.2 MPa \\ 
       \hline
      \end{tabular}
 \caption{Transducer parameters}
  \label{tab:transducer}
\end{center}
\end{table}

The simulation was 9.3 cm deep, 3.9 cm wide, and 2.2 cm in elevation,
which accomodates the footprint of the transducer with a margin to
account for diffraction at the transducer edges. To generate different
transmit sequence the tissue map was translated across this fixed
simulation domain, which is equivalent to translating the
aperture. The spatial step size was set to $\lambda/15$, calculated
relative to the 2 MHz transmit center frequency. The temporal step
size was set to $\Delta x/2.5 c_0$, with $c_0=$ 1540 m/s, which
corresponds to a Courant-Friedrichs-Lewy condition of 0.4.  There were
600 million points in the spatial grid, and the evolution of the
pressure field was calculated over 13,000 time steps, or equivalently
133 $\mu$s. These parameters are summarized in
Table~\ref{tab:simulation}.

\begin{table}[H]
\begin{center}
  \begin{tabular}{|c|c|}
        \hline
       Simulation  & Value \\ 
       Parameter & \\ \hline
       Depth (x) & 9.3 cm\\ 
        Lateral (y) & 3.9 cm \\
        Elevation (z) & 2.2 cm \\ 
        Step size ($\Delta x$) & $\lambda/15$ \\ 
        Step size ($\Delta t$) & $\Delta x/2.5 c_0$ \\ 
        Points & $600\times 10^6$ \\ 
        Time steps & $13\times 10^3$ \\ \hline
      \end{tabular}
  \caption{Simulation parameters}
  \label{tab:simulation}
\end{center}
\end{table}

\section{Results and Discussion}

\subsection{Characteristics of the received signal\label{sec:receive}}

\begin{figure}[H]
  \begin{center}
    \includegraphics[width=0.8\linewidth]{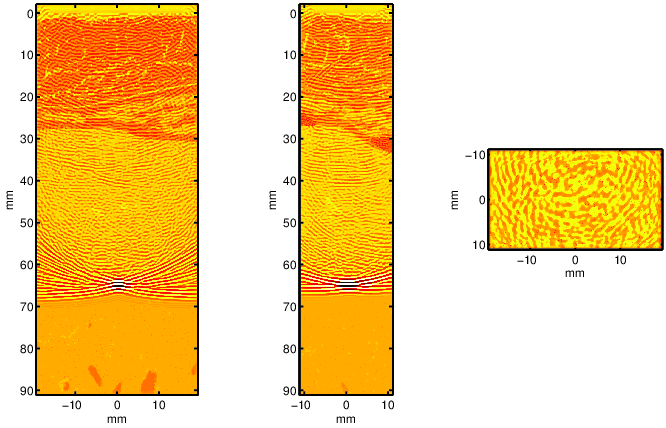}
  \end{center}
  \caption{Snapshot in time of the pressure field on a compressed
    scale as it propagates through a lateral (left) elevation
    (middle) and axial (right) section of tissue.}
  \label{fig:movie}
\end{figure}

Sound reflected from tissue structures and subresolution scatterers
was measured at the grid location corresponding to the transducer
surface, and then used to generate ultrasound images. A time snapshot
of the simulated acoustic field is shown in Fig.~\ref{fig:movie} which
overlays the pressure field on the speed of sound maps from
Fig.~\ref{fig:realization2}. This frame corresponds to the instant in
time when the transmitted pulse has reached the 65 mm focal distance.
The axial section (Fig.~\ref{fig:realization2}, right) is taken at the
transducer surface and the lateral (left) and elevation (middle)
sections are taken at the center of the transducer. Given the large
dynamic range, the pressure field is shown on a compressed scale so
that the low pressure signal in tissue can be seen concurrently with
the transmitted pulse.  A movie of this transmit-receive propagation
event can be found in the online resources.

\begin{figure}[H]
  \begin{center}
     \includegraphics[width=0.39\linewidth]{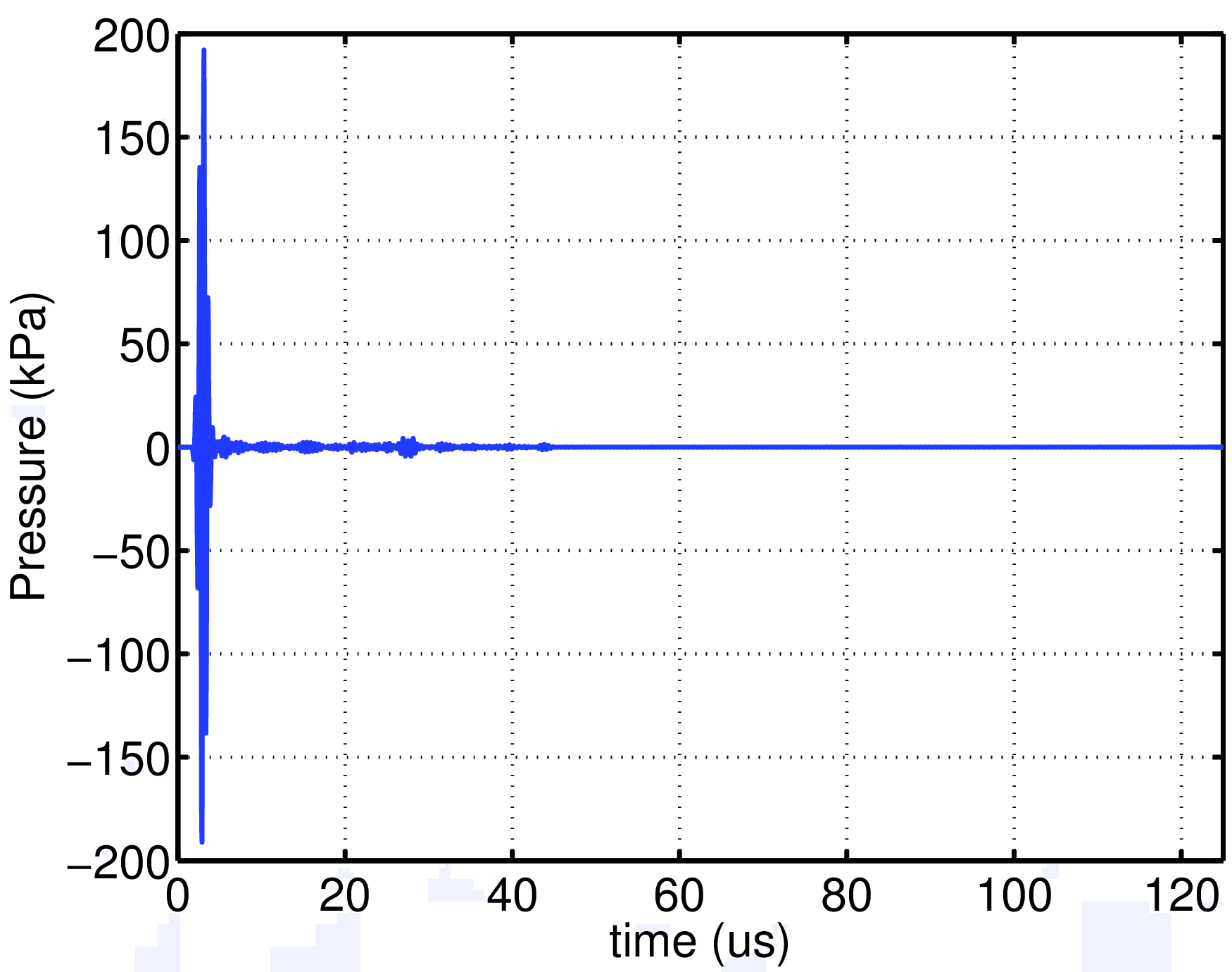}
       \includegraphics[width=0.39\linewidth]{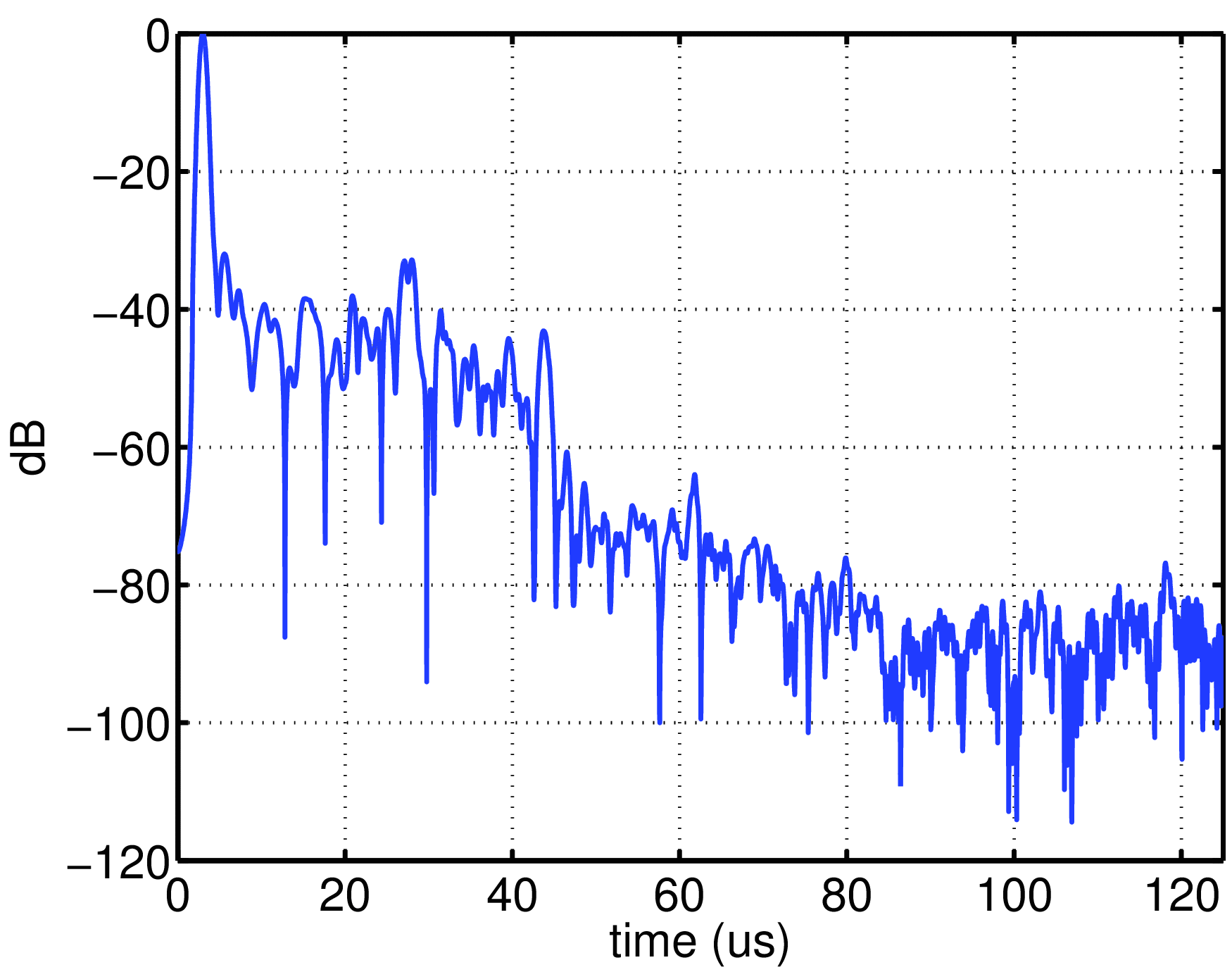}
  \end{center}
  \caption{The pressure as a function of time measured at the center of the virtual transducer on a linear scale (left) and a dB scale (right). }
  \label{fig:timeplot}
\end{figure}

The dynamic range of the acoustic signal captured by the simulation is
illustrated in Fig.~\ref{fig:timeplot}, which shows the signal
measured at the center of the transducer surface as a function of time
on a linear scale (left) and a dB scale (right). On the linear scale
plot the 200 kPa transmitted pulse is clearly visible at at 5
$\mu$s. Following the pulse some backscattered echos are observable up
to 40 $\mu$s however for larger times the echos are too small to
appear on a linear scale. The right plot in Fig.~\ref{fig:timeplot}
shows the envelope detected amplitude on a dB scale. The echo
amplitude gradually decreases as a function of time until it reaches
approximately -80 to -90 dB of the original transmitted
amplitude. This corresponds to the numerical noise floor for the
choice of points per wavelength, the CFL, and the ability of the
perfectly matched layer boundaries to reduce unwanted
reflections. This floor could be lowered, for example, by increasing
the number of points per wavelength.

Unlike a physical transducer the virtual detectors in the simulation
have an almost flat frequency response which is determined by the
comparatively low numerical error in the simulation. An additional
filter could be implemented to model a physical transducer impulse
response. The transducer was sampled at each simulated grid point in
the plane, which corresponds to 201,000 elements. Since the grid size
is $\lambda/15$ and the sampling frequency was set to 100 MHz this
translates into 10 Gb of received data per transmit-receive
event. This is a much finer spatial sampling grid than a physical
array which typically has elements that vary between $\lambda/2$ and
$\lambda$. Averaging over a physical element size can easily be
implemented in post-processing.

\begin{figure}[H]
  \begin{center}
 \includegraphics[width=0.39\linewidth]{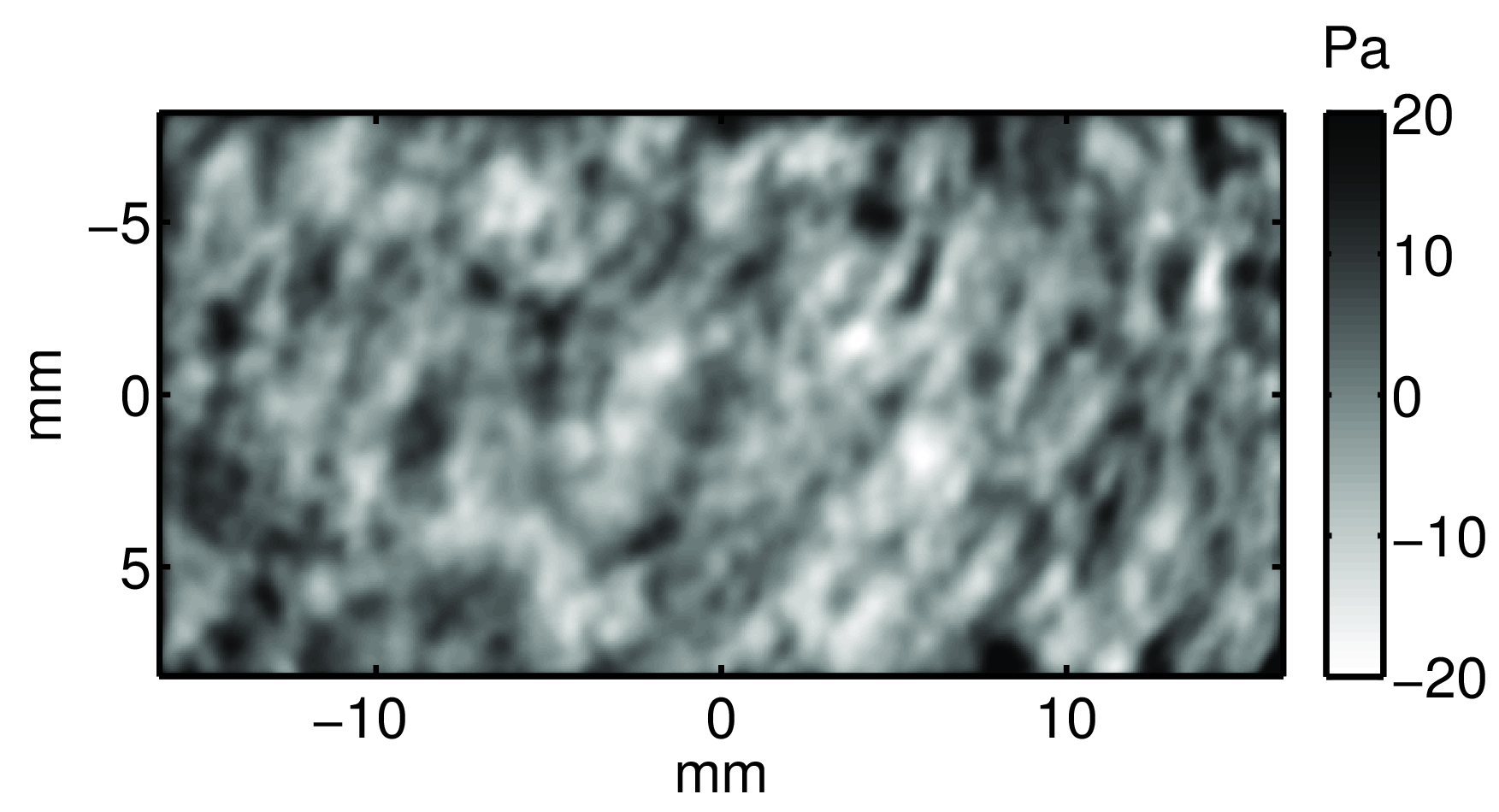}
 \includegraphics[width=0.39\linewidth]{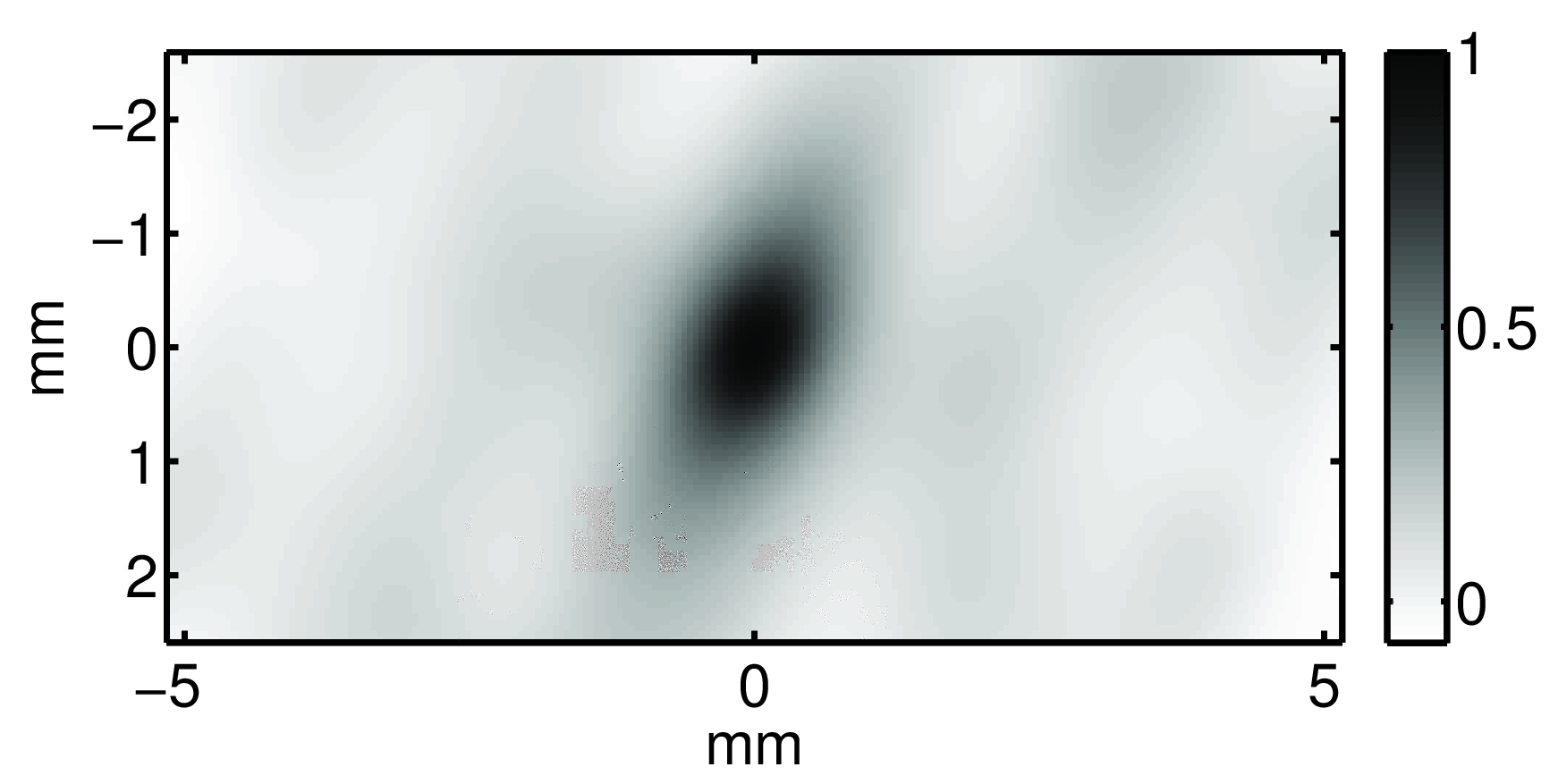}\\
\includegraphics[width=0.42\linewidth]{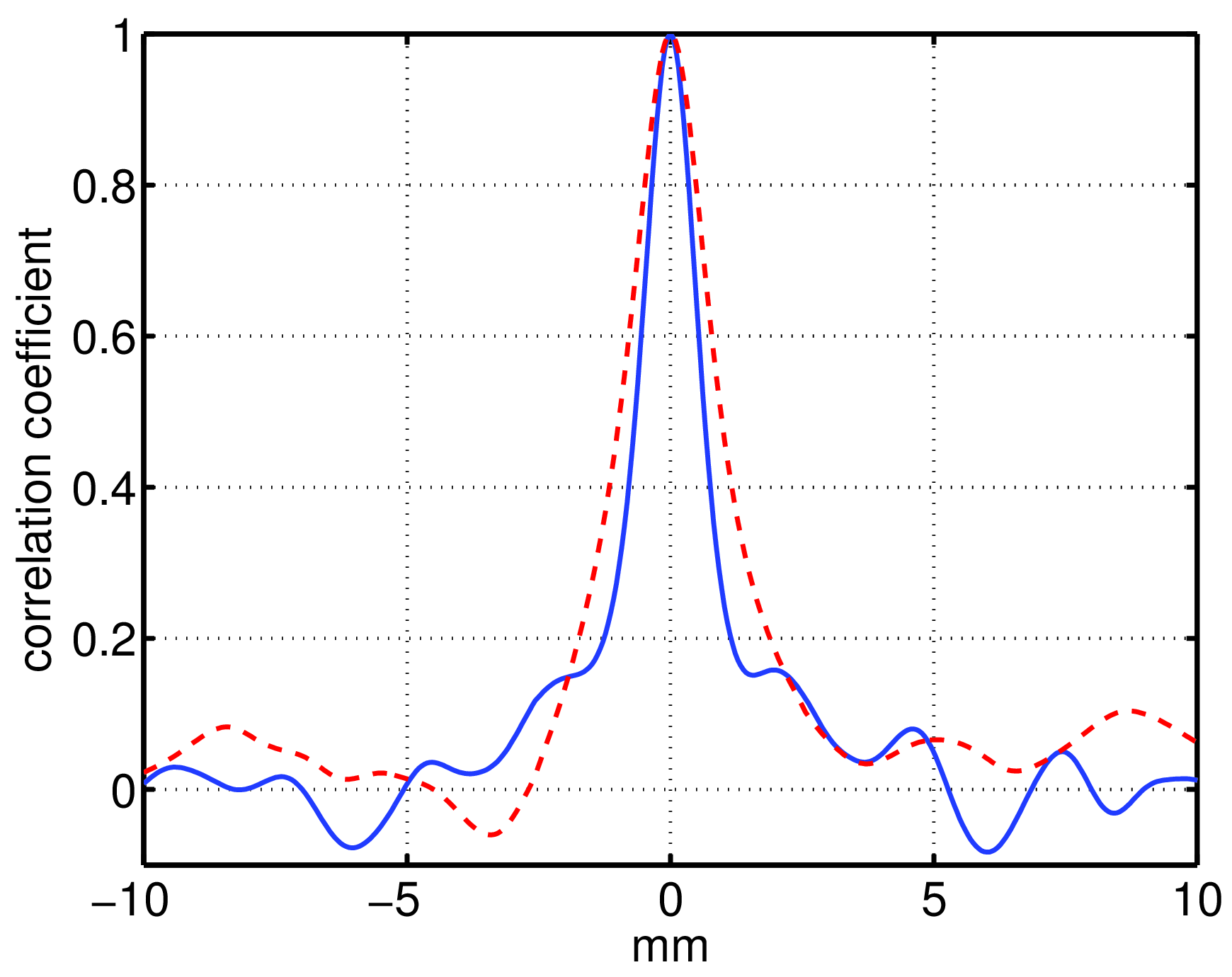}
  \end{center}
 \caption{Signal received at the transducer surface at the focal time
   (top left). Spatial coherence as determined by the auto-correlation
   of the signal received at the transducer surface (top right) and
   the corresponding lateral (solid line) and elevation (dashed line)
   sections of the spatial coherence function (bottom). }
  \label{fig:correlation}
\end{figure}

The ultrasound signal measured at the transducer plane was used to
determine the spatial coherence properties of the signal backscattered
by the human tissue. This parameter characterizes , for example, the
minimum transducer element size required to properly sample the
backscattered pressure field. Note that an equivalent {\it in vivo}
experiment would be difficult to implement since the physical array
element size is typically below the spatial Nyquist frequency. To
calculate the spatial coherence length, first the pressure as a
function of time in the transducer plane was time-gated at the focal
depth and a spherical focusing delay was applied (shown on the top
left of Fig.~\ref{fig:correlation}). The pressure signal was then
autocorrelated to obtain a two dimensional spatial coherence function
(top left plot in Fig.~\ref{fig:correlation}). The lateral and
elevation sections of the transducer correlation are shown on the
bottom right. For this particular body wall realization the full-width
half maximum (FWHM) is $1.71 \lambda$ (1.32 mm) in the lateral
dimension and $2.53\lambda$ (1.95 mm) and in the elevation
dimension. Since the spatial coherence function is anisotropic and has
a preferred orientation the global miniumum of the FWHM is at slight
angle and was measured to be $1.62\lambda$ (1.24 mm). This indicates,
for example, that a uniformly diced two dimensional 2 MHz transducer
array designed for imaging this abdomen would ideally have a $<0.81
\lambda$ ($<$0.62~mm) element size to accurately sample the spatial
variations in the acoustical field.

\subsection{Fundamental and harmonic beams within the body}
One of the advantages of simulations compared to experiments is that
the acoustic field can be estimated in locations where it would not be
possible to scan with a hydrophone. The acoustic field within the
human body, for example, can provide valuable information on the
beamforming characteristics of the ultrasound array. Some general
observations on the beams are made here and a more detailed analysis
on the influence of the beam properties on image quality is given in
Part 2 of this paper.

\begin{figure}[H]
  \begin{center}
    \includegraphics[height=0.4\linewidth]{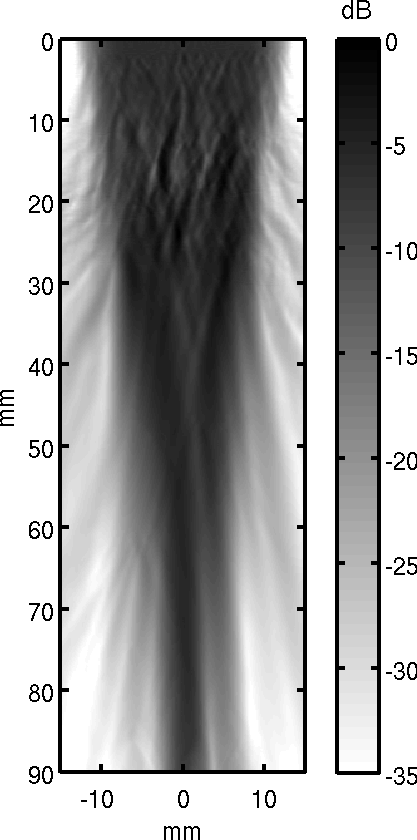}
    \includegraphics[height=0.4\linewidth]{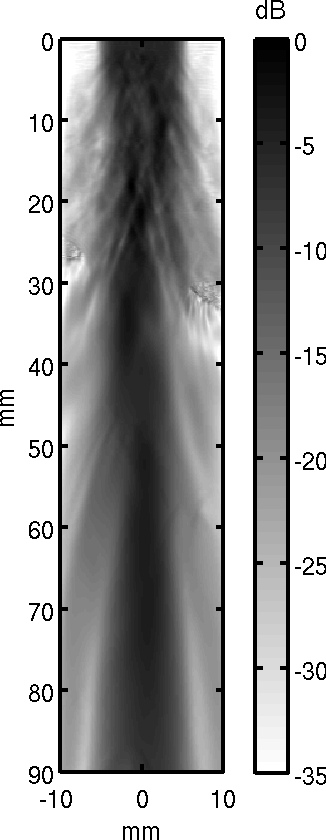}\\
    \includegraphics[height=0.4\linewidth]{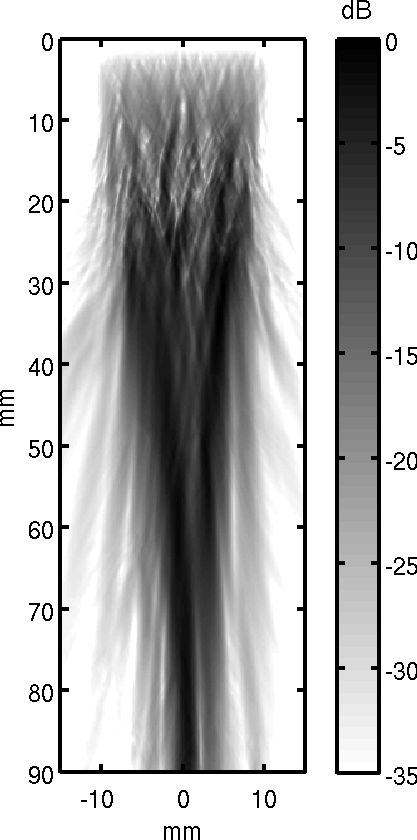}
    \includegraphics[height=0.4\linewidth]{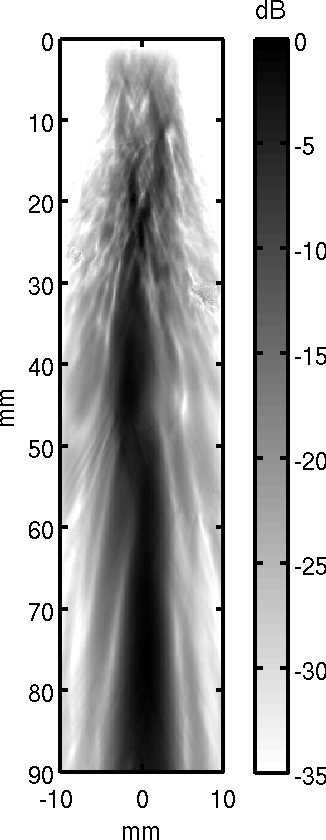}
  \end{center}
  \caption{The intensity in the lateral plane and elevation planes for
    the fundamental frequency (top) and the harmonic frequency
    (bottom). The harmonic beam appears to be fully developed at
    approximately 20 mm of depth, which is before the ribs
    (cf. Fig.~\ref{fig:realization2}).}
  \label{fig:beamplots}
\end{figure}

To investigate the beams at the fundamental and harmonic frequencies
the pressure field was integrated in time for all points in the three
dimensional simulation domain yielding the intensity (shown in
Fig.~\ref{fig:beamplots}). The focal gain is apparent in the lateral
beams (on the left of Fig.~\ref{fig:beamplots}) and the ribs are
visible in the elevation beams (on the right of
Fig.~\ref{fig:beamplots}) as areas of low intensity at 30 and 35 mm
depth. The beam splits into two parts in the lateral dimension and it
deviates significantly from the geometric propagation axis, especially
in the elevation dimension.

\begin{figure}[H]
  \begin{center}

      \includegraphics[width=0.4\linewidth]{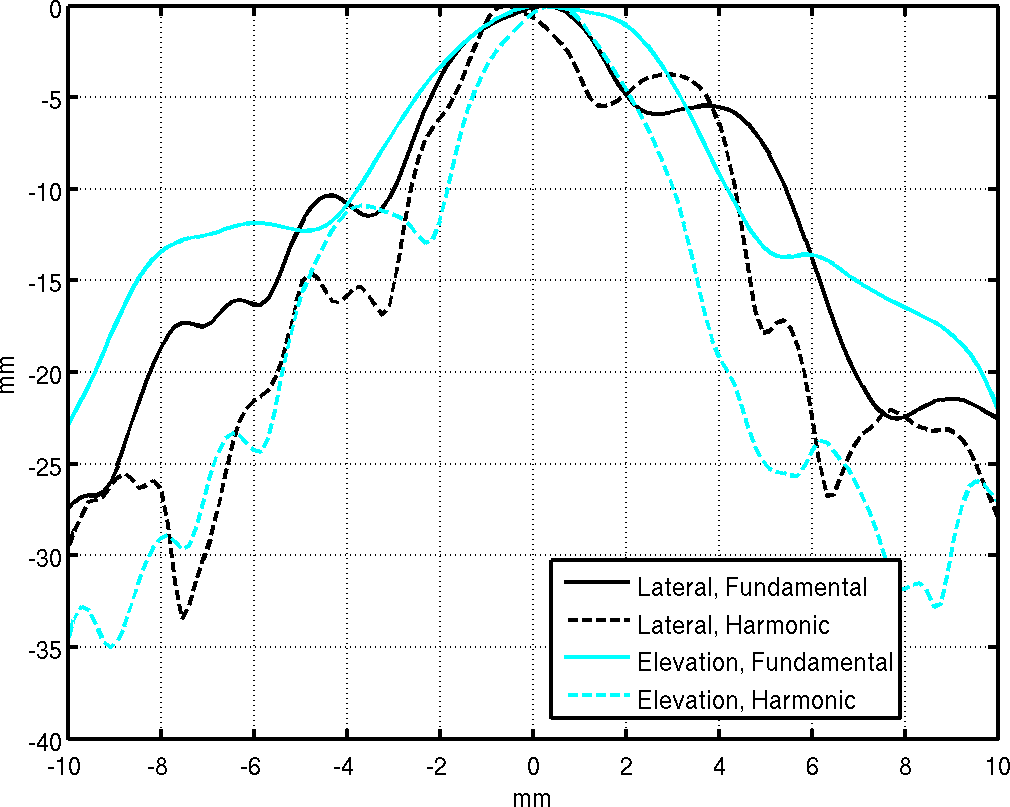}
  \end{center}
 \caption{Beamplots at the focus (65 mm depth) for the beams shown in
   Fig.~\ref{fig:beamplots}. The harmonic beamplots are slightly
   narrower due to the self-focusing effect. However there isn't a
   significant difference between lateral and elevation beamwidth.}
  \label{fig:beamplots2}
\end{figure}

A quantitative comparison between the different beams is shown in
Fig.~\ref{fig:beamplots2} which plots the lateral and elevation
beamplots at the 65 mm focal depth. Even though the transducer has a
2:1 aspect ratio the width of the lateral and elevation beams at the
focus are very similar. This is due to the tissue inhomogeneities
which significantly alter the beam shape throughout its propagation
path (Fig.~\ref{fig:beamplots}). The harmonic beamplots are narrower
than the fundamental beamplots probably due to a self-focusing
effect. Based on the beamplots alone one would therefore hypothesize
that the harmonic images to be better than the fundamental
images. This will be investigated briefly in Sec. \ref{sec:image} and
more thoroughly in Part II of this paper.  Before establishing this
link to image quality, the aberration is described and quantified.




\subsection{Aberration after the ribs}

Phase aberration measures the deviation of a beam from a predetermined
phase profile, such as the ideal focusing profile in a homogeneous
medium. In a heterogeneous medium, or distributed aberrator, such as
the human body, the phase aberration changes as the wave
propagates. Since, in this simulation, the pressure field is known
throughout the body, the phase aberration can be calculated at
different depths.

\begin{figure}[H]
  \begin{center}
    \includegraphics[width=0.39\linewidth]{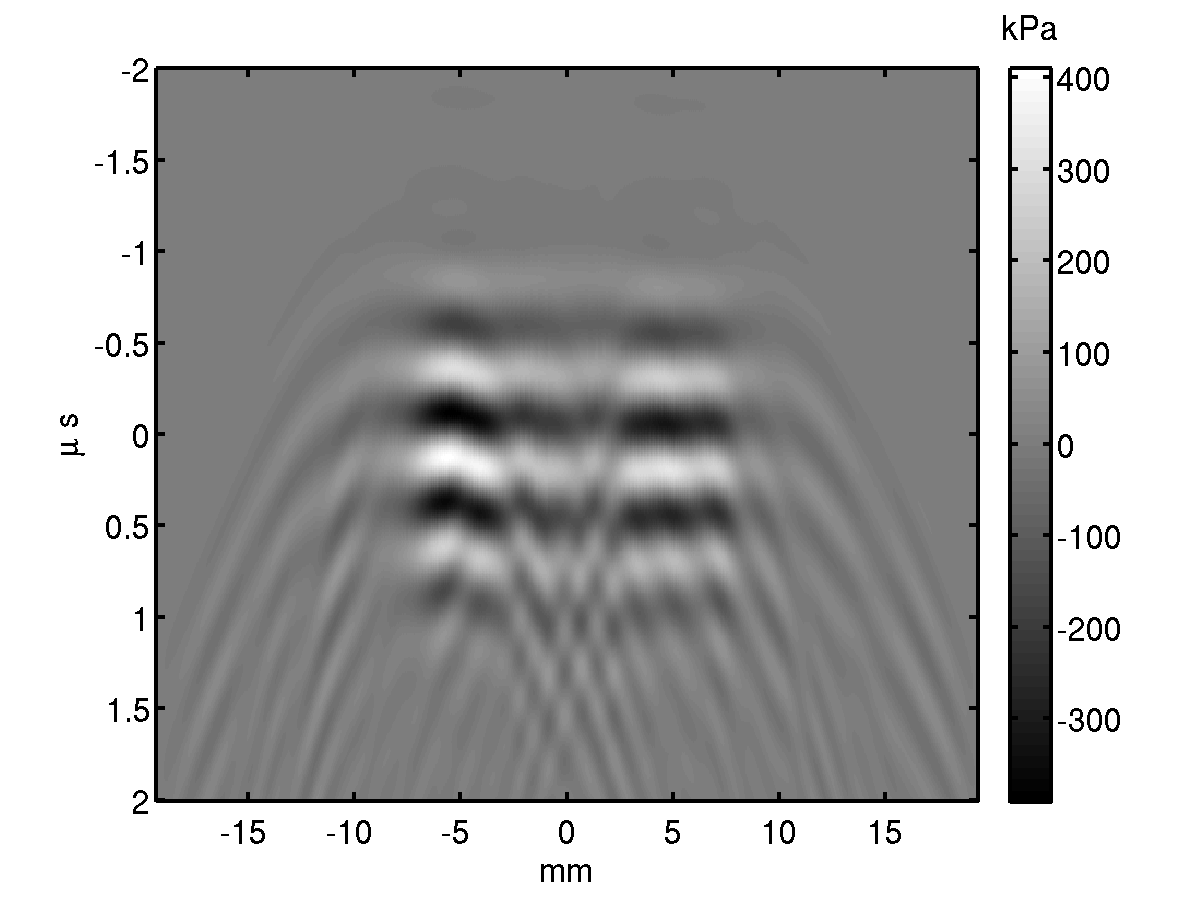} 
    \includegraphics[width=0.39\linewidth]{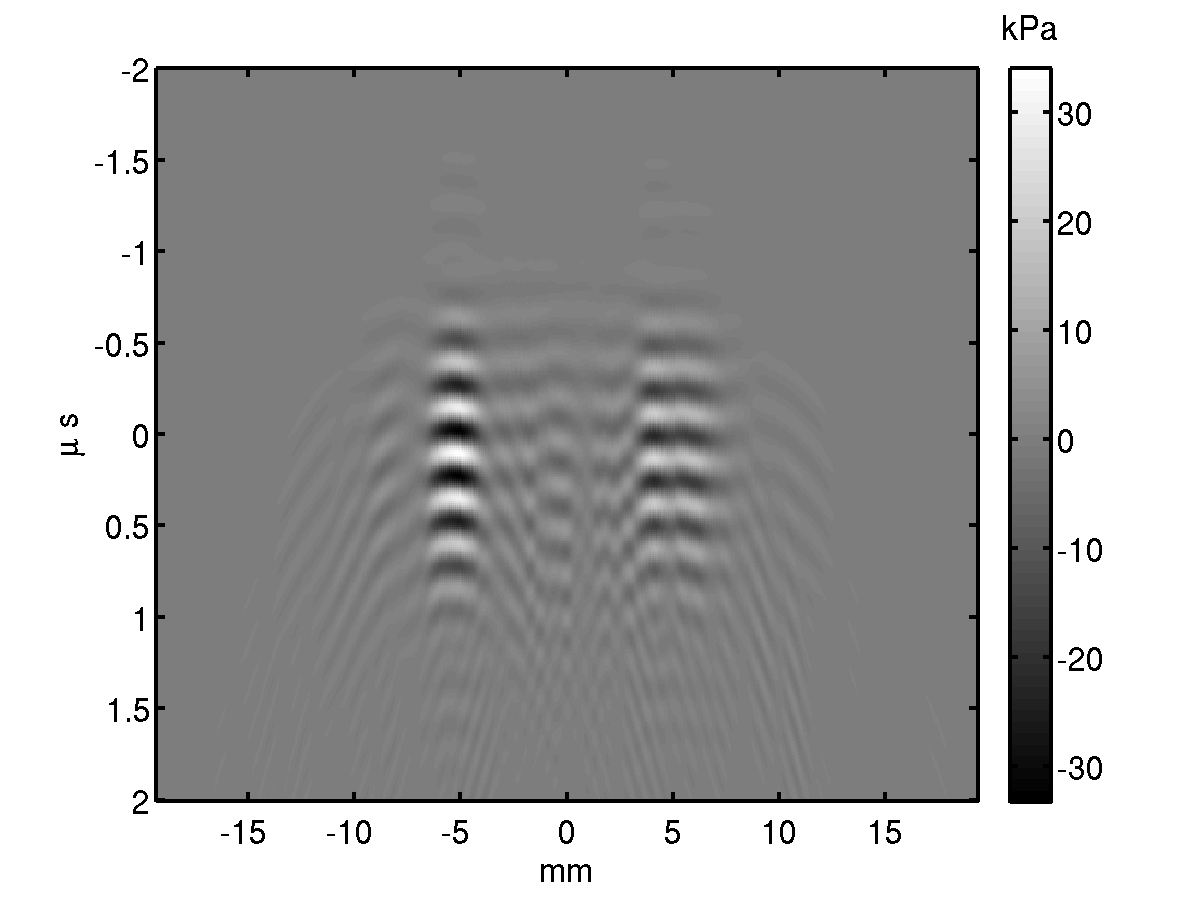}
  \end{center}
  \caption{Received pressure in the lateral axis measured at 3.5 cm
    depth (just after the ribs) at the fundamental frequency (left)
    and harmonic frequency (right). A spherical delay was applied to
    the signals to correct for the focusing profile.}
  \label{fig:aberration}
\end{figure}

Here the the pressure was measured in the lateral and elevation
dimensions at a depth of 35 mm, just after the ribs. The aberration
calculation therefore includes the influence of the near field skin,
connective tissue, fat, and muscle, which are strong aberrators
compared to the relatively homogeneous liver because there is a large
impedance mismatch between different tissue types. To determine the
aberration of the field an inverse spherical delay, representing the
ideal homogeneous medium phase, was applied to the signals. A plot of
this field in the lateral axis is shown in
Fig.~\ref{fig:aberration}. The remaining phase, i.e. any phase
component that isn't constant perpendicular to the propagation axis,
represents the phase error, or phase aberration from the ideal
focusing delay.

The signal was interpolated by a factor of three and then discrete
normalized cross correlation was used to establish discrete estimates
of the time delays. These discrete estimates were then fitted to a
parabola to determine the continuous delay estimates~\cite{pinton2006rapid}.

\begin{table}[H]
  \begin{center}
    \begin{tabular}{|c|c|c|}
      \hline
      & Fundamental &  Harmonic \\ \hline
      lateral &  23.4 ns & 16.2 ns \\
      elevation &  23.5 ns & 12.5 ns \\
      \hline
    \end{tabular}
    \caption{RMS phase aberration at 3.5 cm depth in the lateral and
      elevation dimensions}
    \label{tab:aberration}
  \end{center}
\end{table}

One of the challenges in calculating the phase aberration comes from
areas of diffraction at the beam edges. In these edge diffraction
regions there is significant aberration but very little energy
compared to the central part of the pulse. This is clearly visible in
Fig.~\ref{fig:aberration} where the low energy diffraction edges of
the pulse, beyond $\pm 7$ mm, have deviated significantly from a
planar profile. However these low-energy areas are not particularly
relevant to tissue-generated phase aberration in an imaging context. A
simple threshold was therefore used to eliminate the aberration
estimates where the intensity was less than 10\% of the
maximum. Values above this threshold were used to determine the
root-mean-square (RMS) aberrations. At the fundamental frequency this
aberration is 23.4 ns laterally and 23.5 ns in elevation.  In an {\it
  ex vivo} experiment Hinkelman {\it et
  al.}~\cite{hinkelman1997measurements} determined the deviation of
the arrival time across a mechanically scanned 2-D aperture for
ultrasound propagating through a human chest wall. These
root-mean-square estimates were, on average, 21.3 ns. 

The aberration at the second harmonic frequency is small by slightly
less than a factor of two, 16.2 ns laterally and 12.5 ns in
elevation. These values are summarized in Table~\ref{tab:aberration}.



\subsection{Image analysis \label{sec:image}}

Conventional B-mode images were generated {\it in silico} using the
same propagation-based imaging physics used by an ultrasound
scanner. The 2 MHz pulse was transmitted into the Visible Human
acoustical maps (as described previously in Sec~\ref{sec:transmit}), then the
backscattered signal was received at the transducer location (as
described previously in  Sec~\ref{sec:receive}), and finally conventional delay and sum
beamforming was performed to generate B-mode images.

Beamforming for the two dimensional array was performed in the lateral
and elevation dimensions, across all 201,000 simulation points in the
transducer plane. I.e. each grid point acted as a transducer
element. Since the simulation elements are significantly smaller than
transducer elements the signal could also have been averaged over the
footprint of a single transducer element. Furthermore, data within the
kerf footprint could have been ignored to model a more precise angular
response. However these operations did not make a significant
difference in subsequently presented comparative image quality
analysis and were therefore not performed here.

Dynamic receive focusing was performed in both lateral and elevation
dimensions. A minimum F/\# value of 3 was used to maintain a uniform
speckle pattern. Parallel receive beamforming was used to generate
A-lines across half of a beamwidth (i.e. $1.4\lambda z/2 D$). This
reduced the number of transmit-receive events necessary to generate a
B-mode image to a total of five each with a processing time of 10
hours on a 96 CPU computer cluster. The fundamental and harmonic
compenents of the backscattered signal were calulculated by filtering,
respectively, at the transmit or at two times the transmit frequency
using a standard Gaussian filter.

\begin{figure}[H]
  \centerline{
 \begin{overpic}[width=0.25\linewidth]{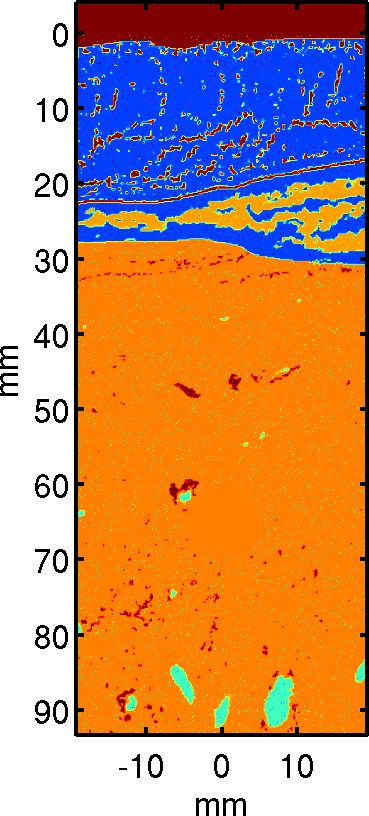}
     \put(71,94){\bf\textcolor{white}{\circle{21}}}
     \put(55,60){\color{white}\line(0,1){165}}
     \put(87,60){\color{white}\line(0,1){165}}
     \put(55,60){\color{white}\line(1,0){32}}
     \put(55,225){\color{white}\line(1,0){32}}
      \put(55,72){\small\bf\textcolor{white}{ anechoic }}
      \put(61,63){\small\bf\textcolor{white}{ lesion }}
\end{overpic}
\raisebox{19mm}{\begin{overpic}[width=0.125\linewidth]{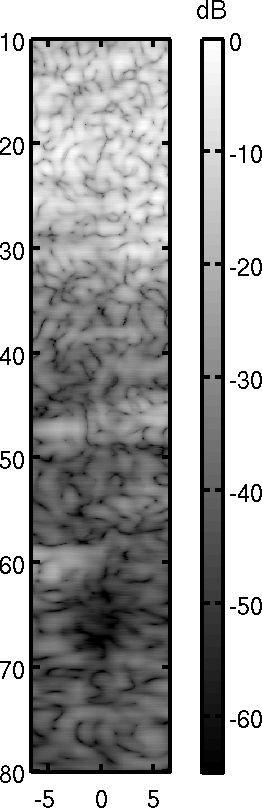}
  \put (16,55) {\textcolor{white}{0.67}}
    \put(5,180){\scriptsize\textcolor{black}{ Fundamental }}
\end{overpic}}
\raisebox{19mm}{\begin{overpic}[width=0.125\linewidth]{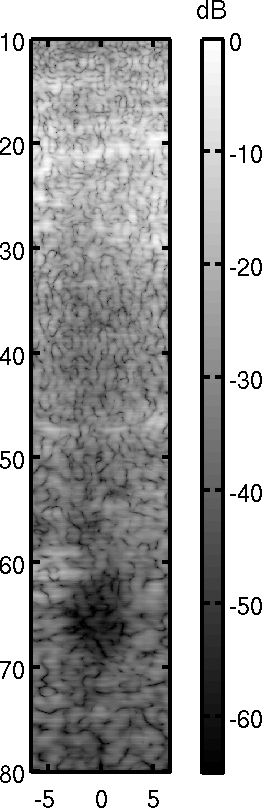}
  \put (16,55) {\textcolor{white}{0.79}}
   \put(7,180){\scriptsize\textcolor{black}{ Harmonic }}
\end{overpic}}
}
 \caption{Harmonic frequency B-mode images calculated with a (from
   left to right) 1D transducer with ribs, 2D transducer without ribs,
   and a 2D transducer with ribs. The corresponding speed of sound map
   is shown on the right.}
  \label{fig:bmode2}
\end{figure}
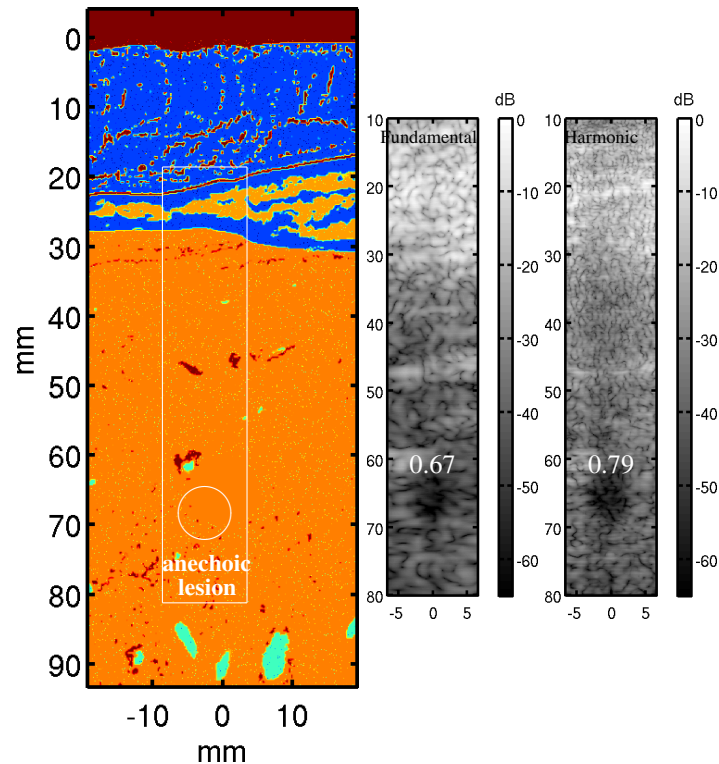

The resulting fundamental and harmonic B-mode images
(Fig.~\ref{fig:bmode2}) clearly show an anechoic lesion that is
visible at the 65 mm focus. The speed of sound map corresponding to
the image location is shown for reference on the left of
Fig.~\ref{fig:bmode2} with the position of the anechoic region lesion
outlined at the focus. As might be expected the contrast to noise
ratio (CNR), which quantifies the lesion detectability, is higher for
the harmonic image (CNR=0.79) than the fundamental image
(CNR=0.67). Features that are observable in the speed of sound map are
also visible in the B-mode images. For example the near field tissue
layers between 0 and 30 mm appear bright and the tissue structures
within the liver at 48 mm, and 60 mm depth are visible.


\section{Summary and conclusion}

In summary the we have described how the optical data from the Visible
Human project can be used to generate anatomically realistic maps of
the acoustical properties human tissue for an intercostal imaging
scenario. A divide and conquer approach that segmented different
regions and then detected specific tissue types within those regions
was used. These optical image to acoustical map transformation
techniques are applicable to other areas of the body. The acoustical
maps can be used to simulate ultrasound imaging, therapeuetic, or
generally any acoustic emission. Here the Fullwave simulation tool was
used to generate highly realistic harmonic and fundamental B-mode
images based on the first principles of propagation and
reflection. Since the acoustical field is known throughout the
simulation domain measurements of the beamplot, phase aberration, and
spatial coherence were also calculated. It was shown that the
fundamental and harmonic beamplots have a similar width, sidelobe
level, and aberration profiles.
In the companion paper the acoustical maps and image characterization
tools established here are used to determine the influence of clutter
in this intercostal imaging scenario. In conclusion, with these
simulations we have generated the first three dimensional, ultrasound
images based on propagation physics through a highly realistic
anatomical model of the human body.

\bibliography{general}
\bibliographystyle{IEEEtran}

\end{document}